\documentstyle[preprint,aps]{revtex}
\begin{document}
\draft
%
%
%
%
\title{
Universal persistent currents in mesoscopic metal rings\\
due to long-range Coulomb interactions}
\author{Peter Kopietz\cite{address}}
\address{
Max-Planck-Institut f\H{u}r Festk\H{o}rperforschung,
Heisenbergstr.1, \\
D-7000 Stuttgart 80, Germany}
\date{\today}
\maketitle
\begin{abstract}
We calculate the average persistent current
in a mesoscopic metal ring threaded by a magnetic flux
in the diffusive regime.
It is shown that the classical electromagnetic energy leads to
a {\it{universal}} average current of the order of $ \alpha e c  / C_{0}$,
where $\alpha$ is the fine structure constant, $-e$
is the charge of the electron, $c$ is the velocity of light,
and $C_{0}$ is the classical capacitance of the ring.
Striking similarities between persistent currents
and universal conductance fluctuations are discovered.
We suggest a simple experiment to test our theory.
\end{abstract}
%
%
%
%
%
\narrowtext
The experimental confirmation\cite{Levy90,Chandrasekhar91} of the existence of
persistent currents in mesoscopic normal-metal rings threaded by a magnetic
flux
$\phi$ has stimulated many recent theoretical
works\cite{Ambegaokar90}-\cite{Kopietz92}.
However, a satisfactory explanation for the experiments has so far not been
found. In all theoretical models the predicted current is
smaller than the experimentally measured one.
In the present work we shall
resolve this discrepancy between theory and experiment.
We show that in the diffusive regime the energy associated with
long-wavelength and low-energy charge fluctuations is determined by classical
charging energies of suitably defined capacitors.
The flux dependence of
these energies yields the dominant contribution
to the persistent current.
Moreover, we show that
the average current is universal precisely in the same
sense as the average variance of the conductance\cite{Lee87}.

This article is divided into two parts. We first study the probability
distribution
of long-wavelength and low-energy density fluctuations in a disordered
metal ring in the diffusive regime.
Combining the insight gained from the first part with
simple semiclassical arguments, we then derive
the universal average current and compare it with experiments.

Consider a thin cylindrical metal ring with
circumference $L$ and cross section $L_{\bot}^2$.
The electrons inside the ring interact with the Coulomb potential,
and are elastically scattered by impurities.
The elastic mean free path $\ell$ is assumed to satisfy
$L \gg \ell \geq L_{\bot}$.
For a proper description of Coulomb effects it is crucial to take the
neutralizing positive background charge density
$\rho_{+} ( {\bf{r}} )$ of the ions into account.
For our purpose it is sufficient to approximate $\rho_{+} ( {\bf{r}} )$
by a constant $\rho_{+}$ for ${\bf{r}}$ inside the ring,
and set $\rho_{+} ( {\bf{r}} ) = 0 $ if ${\bf{r}}$ is outside the ring.
Charge neutrality requires that the average electronic density equals the
density of the positive background charge,
$\overline{ \rho ( {\bf{r}} ) } = \rho_{+} ( {\bf{r}} )$,
where the overbar denotes average over the disorder.
In terms of the exact retarded/advanced Greens functions
$G^{R/A}_{\sigma} ( {\bf{r}} , {\bf{r}}^{\prime} , \epsilon )$
of spin-$\sigma$ electrons, the
true electronic density at temperature $T$ is, for a given realization of the
disorder, given by
	$
	\rho ( {\bf{r}} )  =  \int_{-\infty}^{\infty} d \epsilon f ( {\epsilon } )
	\rho ( {\bf{r}} , \epsilon )
	$,
where $f ( \epsilon ) = [ e^{\epsilon /T } + 1 ]^{-1}$, and
	\begin{equation}
	\rho ( {\bf{r}} , \epsilon )
	 =
	\frac{1}{2 \pi i }
	\sum_{\sigma = \uparrow \downarrow}
	\left[ G^{A}_{\sigma} ( {\bf{r}} , {\bf{r}} , \epsilon )
	- G^{R}_{\sigma} ( {\bf{r}} , {\bf{r}} , \epsilon ) \right]
	\; \; \; .
	\label{eq:rhoedef}
	\end{equation}
We are interested in the probability distribution of the dimensionless random
variables
	\begin{equation}
	\hspace{-5mm}
	N_{k} =
	\int d{\bf{r}} e^{i k \theta }
	\int_{- E_{\tau}}^{E_{\tau} } d \epsilon
	f ( \epsilon ) \rho ( {\bf{r}} , \epsilon )
	\label{eq:Pdef}
	\; \;  , \; \; k = 1,2,\ldots \; ,
	\label{eq:Nkdef}
	\end{equation}
where the ${\bf{r}}$-integration is over the volume
$V = L L_{\bot}^2$ of the ring, and $\tan \theta = y/x$.
We have chosen the coordinate system such that the
ring lies in the $xy$-plane, with the origin at the
center of the ring.
The cutoff energy $E_{\tau}$ is
large compared with the Thouless energy $E_{c} = \hbar {\cal{D}} / L^2$,
but small compared with $\hbar / \tau $.
Here ${\cal{D}}$ is the diffusion coefficient,
and $\tau$ is the elastic lifetime.
The $N_{k}$ are the Fourier components of the electronic density due to
states within an interval of width $2E_{\tau}$ around the Fermi energy.
Below we show that only these states are responsible
for the flux dependence of the energy.
Because $\overline{\rho ( {\bf{r}} , \epsilon )}$
does not depend on ${\bf{r}}$,
the averages $\overline{N_{k}}$ vanish for $k \neq 0$.
More interesting are the variances
	\begin{equation}
	P_{k} = \overline{ \delta N_{k} \delta N_{-k} }
	\label{eq:Pkdef}
	\; \; \; ,
	\end{equation}
where $\delta X = X - \overline{X}$ for any random variable $X$.
We first calculate
	\begin{equation}
	P ( {\bf{r}}, {\bf{r}}^{\prime},  {\epsilon } , {\epsilon }^{\prime} )
	=
	\overline{
	\delta \rho ( {\bf{r}} , \epsilon )
	\delta \rho ( {\bf{r}}^{\prime} , \epsilon^{\prime} )
	}
	\; \; \; .
	\label{eq:Prdef}
	\end{equation}
The $P_{k}$ can then be obtained from
Eqs.\ref{eq:Nkdef} and \ref{eq:Pkdef}.
The essential observation is now that
$P ( {\bf{r}} , {\bf{r}}^{\prime} , \epsilon , \epsilon^{\prime} )$
{\it{is for}} $ | {\bf{r}} - {\bf{r}}^{\prime} | \gg \ell$ {{\it{and}}
$| \epsilon | , | \epsilon^{\prime} | \ll \hbar / \tau$ {\it{universal,
and can be calculated quite reliably
within a model of non-interacting electrons}}\cite{Altshuler86}.
The behavior of this function is determined
by the fact that,
at large distances and low energies,
density fluctuations in a disordered metal are
governed by the diffusion equation.
As long as inelastic processes do not drastically
modify the diffusive behavior,
their effect can be approximately taken into account on a phenomenological
level
by introducing a finite phase breaking energy
in the cooperon- and diffuson propagators
(see Eq.\ref{eq:Cooperon} below)\cite{Altshuler86,Schmid91}.
To justify this more rigorously, let us consider the diagrammatic calculation
of
$P ( {\bf{r}} , {\bf{r}}^{\prime} , \epsilon , \epsilon^{\prime} )$.
A general diagram consists of two closed loops representing the
exact spectral densities $\rho ( {\bf{r}} , \epsilon )$ and
$\rho ( {\bf{r}}^{\prime} , \epsilon^{\prime})$, which are
connected in all possible ways by static impurity lines.
Note that the product of the averages is subtracted in Eq.\ref{eq:Prdef},
so that the loops must be connected together by at least one impurity line.
The dominant flux dependent diagram in the absence
of interactions is shown in Fig.\ref{fig:path}.
In the interacting system, Coulomb lines are not permitted to connect the
loops.
Precisely the same diagrammatic rules are encountered
in the theory of conductance fluctuations\cite{Lee87}.
The similarity of Fig.\ref{fig:path} with the diagrams shown in Fig.5 of
Lee {\it{et al.}}\cite{Lee87} is evident.
Diagrams corresponding to interaction corrections
to $P ( {\bf{r}} , {\bf{r}}^{\prime} , \epsilon , \epsilon^{\prime} )$ can be
obtained
from the diagrams shown in Figs.9 and 10 of Ref.\cite{Lee87}
by simply omitting the current vertices.
The contribution of these diagrams to
$P ({\bf{r}} , {\bf{r}}^{\prime} , \epsilon, \epsilon^{\prime} )$
is negligible {\it{for exactly the
same reasons}} as discussed in Ref.\cite{Lee87}.
The close connection between fluctuations of the conductance
and the density of states has also
been pointed out in Ref.\cite{Altshuler86}.

It follows that a reasonable approximation
for the function
$P ({\bf{r}} , {\bf{r}}^{\prime} , \epsilon, \epsilon^{\prime} )$
in the presence of interactions can be obtained by
performing the calculations for free electrons, and
adding a phase breaking energy $\Gamma$ to the poles of the
cooperons and diffusons.
The dominant flux dependent diagram
is shown in Fig.\ref{fig:path}.
If $\epsilon$ and $\epsilon^{\prime}$ approach the real axis
from opposite sides, this diagram yields\cite{Altshuler86}
	\begin{equation}
	P ( {\bf{r}} , {\bf{r}}^{\prime} , {\epsilon } ,
	{\epsilon}^{\prime}) = 4
	[ {W} ( {\bf{r}} , {\bf{r}}^{\prime}  , \epsilon -
	\epsilon^{\prime}  ) ]^2
	\label{eq:diagram}
	\; \; \; ,
	\end{equation}
where $ [\hbar^2 / ( 2 \pi \nu  \tau^2 )]
{W} ( {\bf{r}} , {\bf{r}}^{\prime} , \epsilon )$ is the usual
cooperon propagator.
($\nu$ is the density of states at the
Fermi energy for one spin species;
the factor of $4$ takes into account the spin degeneracy.)
There exists a similar diagram involving two diffusons, which does not
depend on the flux, and has been ignored in Eq.\ref{eq:diagram}.
It is understood that the $P_{k}$ calculated below contain
only the contribution from the cooperon channel.
The Fourier transform of
${W} ( {\bf{r}} , {\bf{r}}^{\prime} , \epsilon )$
is for $k \ll L / \ell$ and $|\epsilon | \ll E_{\tau} $ given by
	\begin{eqnarray}
	\frac{1}{V}
	\int d{\bf{r}} \int d{\bf{r}}^{\prime} e^{i k ( \theta - \theta^{\prime} )}
	{W} ( {\bf{r}} , {\bf{r}}^{\prime} ,  \epsilon )
	= & &
	\nonumber
	\\
	  \left[ { 4 \pi^2 E_{c} ( k + 2 \varphi )^2 - i \epsilon
	+ \Gamma } \right]^{-1}
	& &
	\label{eq:Cooperon}
	\; \; \; ,
	\end{eqnarray}
where the energy $\Gamma$ models inelastic processes,
such as Coulomb- and electron-phonon interactions\cite{Schmid91,Altshuler86}.
We have defined the dimensionless flux
$\varphi = \phi / \phi_{0}$, where
$\phi_{0} = hc/e$ is the flux quantum.
The real time Fourier transform of
${W} ( {\bf{r}} , {\bf{r}}^{\prime} , \epsilon )$
is precisely the quasi-probability
$\tilde{W}_{t} ( {\bf{r}} , {\bf{r}}^{\prime} )$ introduced
in the semiclassical theory of weak localization by
Chakravarty and Schmid\cite{Chacky86}.
The origin of $\Gamma$ is most transparent in this approach:
inelastic processes destroy the phase coherence
between time reversed paths, and are taken into account
phenomenologically by multiplying the quasi-probabilities
of the non-interacting system by a factor
of $e^{- \Gamma t / \hbar }$.

Given Eqs.\ref{eq:diagram} and \ref{eq:Cooperon}, it is straight forward to
calculate the coefficients $P_{k}$ defined in Eq.\ref{eq:Pkdef}.
We find
	\begin{eqnarray}
	P_{k}  &  = &
	\frac{2}{\pi^2}
	\sum_{m = - \infty}^{\infty}
	\int_{0}^{E_{\tau}} d \epsilon
	\frac{\epsilon \coth \left[ \epsilon / (2 T)  \right]}
	{
	4 \pi^2 E_{c} ( m + 2 \varphi )^2 + \epsilon + \Gamma
	}
	\nonumber
	\\
	&    & \hspace{10mm} \times
	\frac{1}{
	4 \pi^2 E_{c} ( m - k + 2 \varphi )^2 + \epsilon + \Gamma }
	\label{eq:Pkcoop}
	\; \; \; .
	\end{eqnarray}
Evidently the $P_{k}$  are even periodic functions
of $\varphi$ with period $1/2$, so that we may expand
	\begin{equation}
	P_{k}
	 =  \frac{P^{(0)}_{k}}{2} + \sum_{m=1}^{\infty}
	P^{(m)}_{k}
	\cos{ \left( 4 \pi m \varphi \right) }
	\label{eq:Pkfour}
	\; \; \; .
	\end{equation}
After a simple contour integration we obtain
	\begin{eqnarray}
	P^{(m)}_{k} = \frac{1}{\pi^2}
	\int_{0}^{  {E_{\tau} }/{ E_{c}}  } dx
	\frac{x^{1/2}} {( \pi k )^2 + x + \Gamma / E_{c}}
	\hspace{10mm}
	& &
	\nonumber
	\\
	\times
	\exp \left[ {-m ({x + \Gamma / E_{c} )^{1/2} } } \right]
	\coth{ \left[ x E_{c} / (2 T) \right] }
	& &
	\; .
	\label{eq:Pkcont}
	\end{eqnarray}
Note that the $P_{k}^{(0)}$ are
for $( \pi k )^2 + \Gamma / E_{c} \ll E_{\tau} / E_{c}$
proportional to $({ E_{\tau}  / E_{c} })^{1/2} = O( L / \ell)$.
The cutoff dependence of all non-zero
Fourier coefficients $P_{k}^{(m)} , m \geq 1$, is exponentially small for large
$L / \ell$, so that we may let $E_{\tau} \rightarrow \infty$.
In the limit $T, \Gamma \ll E_{c}$, the $P_{k}^{(m)}$
are for $m \geq 1$ universal numbers, determined by the
shape of the system.
It is important to stress that Eq.\ref{eq:Pkcont}
includes the effect of inelastic processes, which are, however,
negligible if the effective inelastic energy $\Gamma$
is small compared with $E_{c}$.

We now develop our semiclassical theory of persistent currents.
Consider first the charge distribution on the ring
associated with the first harmonic $N_{1}$ of the density fluctuation.
Although $\overline{N_{1}} = 0$,
the {\it{typical}} excess charge on one side of the ring is
$ e ({ \overline{ N_{1}^2} })^{1/2}  =  e ({P_{1}})^{1/2} \propto e  ({L/ \ell
})^{1/2}$.
Because by assumption
$( L / \ell )^{1/2}  \gg 1$, the associated electrostatic energy can be
calculated
classically, and is given by
$e^2 P_{1} / (2 C_{1})$, where $C_{1}$ is the capacitance
of a thin ring consisting of two oppositely charged halfs.
The contribution of modes with wavelengths large compared with $\ell$
to the disorder averaged Hartree energy is simply
the sum of the corresponding classical charging energies,
	\begin{equation}
	\overline{\Omega}_{\phi}  =  \sum_{k=1}^{\Lambda}
	\frac{e^2}{2 C_{k}} P_{k}
	\; \; \; ,
	\label{eq:Hartree}
	\end{equation}
where the integer $\Lambda $ is chosen such that $1 \ll  \Lambda  \ll L /
\ell$,
and the generalized capacitances $C_{k}$ are
	\begin{equation}
	\frac{1}{C_{k}} = \frac{1}{V^2} \int d{\bf{r}}
	\int d{\bf{r}}^{\prime}
	\frac{ e^{i    k ( \theta - \theta^{\prime} )}  }
	{ | {\bf{r}} - {\bf{r}}^{\prime} | }
	\; \; \; .
	\label{eq:capacitance}
	\end{equation}
The sum in Eq.\ref{eq:Hartree} does not include
the term $k=0$, because
$\int d {\bf{r}} \delta \rho ( {\bf{r}} ) = 0$ by charge neutrality.
For $k \ll L/L_{\bot}$ we may approximate to leading logarithmic order
$C_{k} \approx C_{0} \approx L / [ 2 \ln ( L / L_{\bot} )]$,
which is the classical capacitance of the ring.
The identification of Eq.\ref{eq:capacitance} with
classical capacitances is only valid for $L_{\bot} \ll L$.
The possibility that charging energies could lead to large
persistent currents has already been noticed by
Imry and Altshuler\cite{Altshuler92}.
Although they examined only the zero mode $C_{0}$,
Imry speculated that local charge fluctuations
in a globally neutral system might be the key to understand persistent
currents.

Eq.\ref{eq:Hartree} can also be derived formally.
A rigorous theorem due to Hohenberg and Kohn\cite{Hohenberg64}
tells us that the {\it{exact}} Hartree energy
of all positive and negative charges
is a functional of the true electronic density $\rho ( {\bf{r}} )$,
	\begin{equation}
	\Omega = \frac{e^2}{2}
	\int d{\bf{r}} \int d{\bf{r}}^{\prime}
	\frac{ \left[ \rho ( {\bf{r}} ) - \rho_{+}  \right]
	\left[ \rho ( {\bf{r}}^{\prime} ) - \rho_{+}  \right] }
	{ | {\bf{r}} - {\bf{r}}^{\prime} | }
	\; \; .
	\label{eq:emenergy}
	\end{equation}
Although Eq.\ref{eq:emenergy} contains explicitly only the
bare Coulomb interaction,
all higher order interaction processes are taken into account
by demanding that $\rho ( {\bf{r}} )$ is,
for a given disorder potential, the true electronic density
of the interacting many-body system.
In an infinite homogeneous system without disorder Eq.\ref{eq:emenergy}
vanishes,
but for a mesoscopic disordered ring $\Omega > 0$.
Charge neutrality implies $\rho_{+} = \overline{\rho ( {\bf{r}} ) }$, so that
the average $\overline{\Omega}$ is
{\it{uniquely determined by the variance of the density}}.
Note the essential role of the positive background charge
to subtract the product of the average densities.
To isolate the flux dependent part of $\overline{\Omega}$,
it is sufficient to keep only the
contribution from modes with wavelengths large compared with $\ell$, involving
states close to the Fermi energy.
According to Eq.\ref{eq:Pkcoop} only these modes are sensitive to the cooperon
pole.
Using the definitions \ref{eq:Nkdef}, \ref{eq:Pkdef}, and \ref{eq:capacitance},
we arrive precisely at Eq.\ref{eq:Hartree}.
The physical reason for the absence of screening in Eq.\ref{eq:Hartree}
is that the random potential creates in the diffusive regime
typical configurations where a large number of excess electrons
is localized on segments of the ring.
Therefore the ring behaves as if it were a classical capacitor.

The sequence of bubble diagrams retained in the
random-phase approximation
does not contribute to the variances $P_{k}$.
In Fig.\ref{fig:screen} we show the first bubble correction to
Fig.\ref{fig:path}.
Because the poles of both Greens
functions associated with the loop
labeled by $\epsilon^{\prime}$ lie in the same half-plane,
the ${\bf{r}}_{5}$-integration leads to the vanishing of the diagram.
Furthermore, if the inner left loop in Fig.\ref{fig:screen}
is not connected by impurity lines to the other loops,
the resulting diagram is exactly cancelled by another diagram
of the same form, where the inner left loop represents the density
of the positive background charge.
This diagram arises from the term in the Hamiltonian that represents
the interaction between electronic and ionic densities.
For precisely the same reasons diagrams of this type
can be ignored in the theory of conductance fluctuations\cite{Lee87}.
An essential difference between our
approach and Ref.\cite{Ambegaokar90} is that
we have included the effect of the
positive background from the very beginning
in a non-perturbative way.
The importance of charge neutrality
has also been emphasized by Schmid\cite{Schmid91}.

{}From Eqs.\ref{eq:Pkfour},\ref{eq:Pkcont} and \ref{eq:Hartree} we obtain
the Hartree contribution to the average persistent current,
	\begin{equation}
	\overline{I} =
	 - c \partial{\overline{\Omega}_{\phi}} / \partial \phi
	=
	\sum_{m=1}^{\infty} I^{(m)} \sin ( 4 \pi m \varphi )
	\; \; \; ,
	\label{eq:Ires}
	\end{equation}
with
	\begin{eqnarray}
	I^{(m)} = \frac{ 2 \alpha}{\pi^2 }
	\sum_{k=1}^{\infty} \frac{ec}{C_{k}}
	\int_{0}^{\infty} dx
	\frac{x^2}
	{   ( \pi m k )^2 + x^2  + \gamma_{m} }
	 & &
	\nonumber
	\\
	\times
	\exp \left[ - ( x^2 + \gamma_{m} )^{1/2}  \right]
	\coth{ \left[ x^2  \beta_{m} / 2  \right] }
	& &
	\; \; ,
	\label{eq:Im}
	\end{eqnarray}
where $\beta_{m} = E_{c} / (m^2 T)$, $\gamma_{m} = m^2 \Gamma / E_{c}$, and
$\alpha = e^2 / ( \hbar c ) \approx \frac{1}{137}$ is the fine structure
constant.
The summand in Eq.\ref{eq:Im} vanishes as $k^{-2}$ for large $k$, so that
the value of the sum is dominated by the first few terms,
and we have set $\Lambda = \infty$ in Eq.\ref{eq:Hartree}.
Eq.\ref{eq:Im} is the main result of this work.
In the limit $T , \Gamma \ll E_{c}$, the current
is {\it{independent of any microscopic details of the metal}},
and depends only on fundamental physical constants and
generalized capacitances $C_{k}$, which are completely determined
by the large-distance geometry of the system.
The average variance of the conductance
in the diffusive regime is universal in precisely
the same sense.

Note that $ \Gamma \raisebox{-0.5ex}{$\; \stackrel{<}{\sim} \;$} E_{c}$
in the experiments\cite{Levy90,Chandrasekhar91}.
For $\Gamma \gg E_{c}$ the current would be exponentially small.
To estimate the magnitude of $I^{(m)}$ for $\gamma_{m} \ll 1$,
we may neglect $x^2$ compared with $( \pi m k)^2 $ in the
denominator of Eq.\ref{eq:Im},
because the factor $e^{-x}$ cuts off the integration at $x= O(1)$.
In this approximation we obtain
	\begin{equation}
	\frac{I^{(m)}}{I_{F}}
	\approx
	- \frac{8 \zeta (2) \ln ( L / L_{\bot} )}{\pi^4 m^2}
	\frac{ c}{137 v_{F}}
	g \left( \beta_{m}  \right)
	\; \; \; ,
	\label{eq:Im3}
	\end{equation}
where $I_{F} = (-e) v_{F} / L $, $\zeta (2) \approx 1.64$, and
	\begin{equation}
	g ( \beta )  =  \frac{1}{2} \int_{0}^{\infty} dx x^2 e^{-x}
	\coth{ \left[ x^2  \beta /2  \right] }
	\label{eq:fmdef}
	\; \; \; .
	\end{equation}
For  $T \ll E_{c} / m^2$ we have $ g ( \beta_{m} ) \approx g ( \infty ) = 1$.
At finite temperature, the relevant energy scale where the current
begins to deviate from its value at $T=0$ is $E_{c}$.
The temperature dependence of Eq.\ref{eq:Im3} is
in agreement with Ambegaokar and Eckern\cite{Ambegaokar90}. As pointed out
in Ref.\cite{Ambegaokar90}, this temperature dependence
seems to fit the experiment\cite{Levy90}.
Moreover, $I^{(m)} \propto m^{-2}$, so that also
the shape of our current agrees with Ref.\cite{Ambegaokar90}.
However, our current is two orders of magnitude larger than the
current calculated in all previous works.
Assuming $L / L_{\bot} = 100$
and $T, \Gamma \ll E_{c}$,
we obtain for the first harmonic
$I^{(1)} \approx - 0.9 \times I_{F}$.
The amazing result is that $I^{(1)}$
has the same order of magnitude as
in the ballistic regime\cite{Kopietz92}.

Our theory can be tested experimentally by embedding
the ring in an insulator
with dielectric constant $\epsilon_{r}$.
Because capacitances increase linearly with $\epsilon_{r}$,
the current should then be reduced by a factor of $1/ \epsilon_{r}$.
The currents measured by Chandrasekhar {\it{et al.}}\cite{Chandrasekhar91}
in three different isolated gold rings
are of order $I_{F}$. Although a disorder ensemble is not appropriate
for a description of this experiment,
the fact that the magnitude of the measured current agrees with our calculation
might indicate that the physics underlying our theory
is also relevant for the
experiment of Ref.\cite{Chandrasekhar91}.
L\'{e}vy {\it{et al.}}\cite{Levy90} have measured an average current of
the order of $ 0.003 \times  I_{F}$.
The system studied in Ref.\cite{Levy90} consists of an array of
$10^{7}$ copper rings with circumference
$L \approx 2.2 \mu m$, occupying a $7 mm^2$ area on a sapphire substrate.
Note that the dielectric constant of sapphire is $\epsilon_{r} \approx 10$,
while in the experiment of Ref.\cite{Chandrasekhar91}
the substrate consists of $Si$, which has $\epsilon_{r} \approx 3$.
Taking the large
dielectric constant into account and assuming
$T = \Gamma = 0$, our theory predicts for the experiment of L\'{e}vy
{\it{et al.}} $ \overline{I} \approx - 0.1 \times I_{F}$,
a factor of $30$ larger than experimentally measured.
This discrepancy can have many explanations.
{}From Eq.\ref{eq:Im} it is clear that an effective
$\Gamma \approx E_{c}$ can easily explain a
reduction of $\overline{I}$ by an order of magnitude.
A second possibility is that
inter-ring interactions and capacitances between rings are important.
Note that in the experimental configuration of Ref.\cite{Levy90}
the average distance between neighboring rings
is smaller than $L$, so that
the Coulomb interaction between rings is not negligible.

Our theory is based on the Hartree energy.
Quantum mechanical exchange- and correlation effects have been ignored.
We have explicitly verified that the current obtained
from the first order exchange correction is
at least a factor of $\ell / L$ smaller than the contribution of the Hartree
term.
Because the mechanism leading to the large current
has a simple semiclassical interpretation,
we believe that quantum mechanical correlation effects
do not essentially modify our result.

In this paper we have
developed a semiclassical theory of persistent currents
in the diffusive regime. The only quantum mechanical
ingredient in our theory are the time reversed paths of
weak localization. The dominant contribution to the average persistent current
is universal, and can be derived from the
classical electromagnetic energy of the system.

This work would have never been completed without
the advise and criticism of K. B. Efetov.
I have also profited from discussions with S. Iida, L. P. L\'{e}vy, and A.
Benoit.

%

\newpage

\noindent
{\Large \bf Figure Captions}
\vspace{7mm}
\hspace{\parindent}

\newcounter{fig}

\begin{list}%
{\bf  Fig.\arabic{fig}:}
{\usecounter{fig} \setlength{\rightmargin}{\leftmargin}}

\item
{Dominant flux dependent diagram determining
$P ( {\bf{r}} , {\bf{r}}^{\prime} , \epsilon , \epsilon ^{\prime} )$.
Solid lines with arrow
denote disorder averaged Greens functions,
dashed lines denote impurity scattering, and the maximally crossed
ladders define cooperons.
\label{fig:path}
}

\item
{
First order bubble correction to Fig.\ref{fig:path}.
The wavy line is the bare Coulomb interaction.
\label{fig:screen}
}
\end{list}

\end{document}